\documentstyle[epsf]{mn}
\newif\ifAMStwofonts
\def\gtorder{\mathrel{\raise.3ex\hbox{$>$}\mkern-14mu
             \lower0.6ex\hbox{$\sim$}}}
\def\ltorder{\mathrel{\raise.3ex\hbox{$<$}\mkern-14mu
             \lower0.6ex\hbox{$\sim$}}}

\ifoldfss
  \ifCUPmtlplainloaded \else
    \NewTextAlphabet{textbfit} {cmbxti10} {}
    \NewTextAlphabet{textbfss} {cmssbx10} {}
    \NewMathAlphabet{mathbfit} {cmbxti10} {} % for math mode
    \NewMathAlphabet{mathbfss} {cmssbx10} {} %  "   "    "
  \fi
  \ifAMStwofonts
    \ifCUPmtlplainloaded \else
      \NewSymbolFont{upmath} {eurm10}
      \NewSymbolFont{AMSa} {msam10}
      \NewMathSymbol{\upi}     {0}{upmath}{19}
      \NewMathSymbol{\umu}     {0}{upmath}{16}
      \NewMathSymbol{\upartial}{0}{upmath}{40}
      \NewMathSymbol{\leqslant}{3}{AMSa}{36}
      \NewMathSymbol{\geqslant}{3}{AMSa}{3E}

      \let\leq=\leqslant 
      \let\geq=\geqslant 
    \fi
  \fi
\fi % End of OFSS

\ifnfssone
  \newmathalphabet{\mathit}
  \addtoversion{normal}{\mathit}{cmr}{m}{it}
  \addtoversion{bold}{\mathit}{cmr}{bx}{it}
  \newmathalphabet{\mathbfit} % math mode version of \textbfit{..}
  \addtoversion{normal}{\mathbfit}{cmr}{bx}{it}
  \addtoversion{bold}{\mathbfit}{cmr}{bx}{it}
  \newmathalphabet{\mathbfss} % math mode version of \textbfss{..}
  \addtoversion{normal}{\mathbfss}{cmss}{bx}{n}
  \addtoversion{bold}{\mathbfss}{cmss}{bx}{n}
  \ifAMStwofonts
    \ifCUPmtlplainloaded \else
      \UseAMStwoboldmath
      \makeatletter
      \new@mathgroup\upmath@group
      \define@mathgroup\mv@normal\upmath@group{eur}{m}{n}
      \define@mathgroup\mv@bold\upmath@group{eur}{b}{n}
      \edef\UPM{\hexnumber\upmath@group}
      \new@mathgroup\amsa@group
      \define@mathgroup\mv@normal\amsa@group{msa}{m}{n}
      \define@mathgroup\mv@bold\amsa@group{msa}{m}{n}
      \edef\AMSa{\hexnumber\amsa@group}
      \makeatother
      \mathchardef\upi="0\UPM19
      \mathchardef\umu="0\UPM16
      \mathchardef\upartial="0\UPM40
      \mathchardef\leqslant="3\AMSa36
      \mathchardef\geqslant="3\AMSa3E

      \let\leq=\leqslant 
      \let\geq=\geqslant 
    \fi
  \fi
\fi % End of NFSS release 1

\ifnfsstwo
  \DeclareMathAlphabet{\mathbfit}{OT1}{cmr}{bx}{it}
  \SetMathAlphabet\mathbfit{bold}{OT1}{cmr}{bx}{it}
  \DeclareMathAlphabet{\mathbfss}{OT1}{cmss}{bx}{n}
  \SetMathAlphabet\mathbfss{bold}{OT1}{cmss}{bx}{n}
  \ifAMStwofonts
    \ifCUPmtlplainloaded \else
      \DeclareSymbolFont{UPM}{U}{eur}{m}{n}
      \SetSymbolFont{UPM}{bold}{U}{eur}{b}{n}
      \DeclareSymbolFont{AMSa}{U}{msa}{m}{n}
      \DeclareMathSymbol{\upi}{0}{UPM}{"19}
      \DeclareMathSymbol{\umu}{0}{UPM}{"16}
      \DeclareMathSymbol{\upartial}{0}{UPM}{"40}
      \DeclareMathSymbol{\leqslant}{3}{AMSa}{"36}
      \DeclareMathSymbol{\geqslant}{3}{AMSa}{"3E}

      \let\leq=\leqslant 
      \let\geq=\geqslant 
    \fi
  \fi
\fi % End of NFSS release 2

\ifCUPmtlplainloaded \else
  \ifAMStwofonts \else % If no AMS fonts
    \def\upi{\pi}
    \def\umu{\mu}
    \def\upartial{\partial}
  \fi
\fi

\title{A Survey for Large Separation Lensed FIRST Quasars, II. Magnification Bias and Redshift Distribution}

\author[E.O. Ofek et al.]
       {Eran O. Ofek\thanks{e-mail: eran@wise.tau.ac.il}$^{1,2}$,  Hans-Walter Rix$^{2}$, Dan Maoz$^{1}$, Francisco Prada$^{3}$ \\
$^{1}$ School of Physics and Astronomy and Wise Observatory, Tel Aviv University, Tel Aviv 69978, Israel \\
$^{2}$ Max-Planck-Institut f\"{u}r Astronomie, K\"{o}nigstuhl 17, D-69117 Heidelberg, Germany \\
$^{3}$ Isaac Newton Group, Apartado de Correos 368, Santa Cruz de La Palma, 38750 Tenerife, Canary Islands, Spain}
\date{Accepted ?
      Received ?
      in original form ?}

\begin{document}

\maketitle

\begin{abstract}
The statistics of large-separation gravitational lensing
are a powerful tool to
probe mass distributions on the scale of galaxy clusters.
In this paper we refine the analysis of
our survey for large-separation ($>5''$)
lensed FIRST quasars (Ofek~et~al. 2001)
by estimating the magnification bias and the source redshift
distribution.
Finding no large separation lens
among 8000 likely quasars in that sample, 
implies an
upper bound on the lensed fraction of $3.7\times10^{-4}$ at $95\%$ CL.
From a published deep $1.4$~GHz radio survey of the Hubble Deep Field,
and corresponding
optical searches for faint quasars,
we calculate a lower limit to the `double flux magnification bias'
affecting our radio-optically selected sample, of $B\geq1.1$.
From the four-colour information in the
SDSS Early Data Release,
we calculate the photometric redshift distribution
of a sample of FIRST quasar candidates
and compare it with the redshift distribution
from the FIRST Bright Quasar Survey.
We find that the median redshift of the quasars in
our sample is about 1.4.
With these new results, we find
that for all plausible cosmologies,
the absence of lensed quasars in our survey is
consistent with a model based on an empirical, non-evolving,
cluster mass function,
where clusters
are represented by
singular isothermal spheres.
On the other hand, 
comparison of our results to the lensing predictions of
published $N$-body-ray-tracing simulations
(Wambsganss et al. 1995, 1998)
rejects
the COBE normalised $\Omega_{0}=1$ CDM model
at $99.9\%$ confidence.

\end{abstract}


\section{Introduction}

The statistics of gravitational lensing can provide a powerful probe
of the geometry and the mass content of the universe
out to large redshifts (e.g. Refsdal 1964; Press \& Gunn 1973; 
Turner, Ostriker, \& Gott 1984).
Lensed quasar statistics
in the angular image separation 
ranges expected from galaxy-mass lenses have been probed
by several surveys
(e.g., Maoz et al. 1993).
However, systematic searches for lensed quasars with multiple
image separations of
$5''\ltorder\Delta\theta\ltorder30''$
expected from groups or clusters
of galaxies
have been carried out
only recently
(e.g., Maoz et al. 1997; Marlow et al. 1998;
Phillips et al. 2001a;
Phillips, Browne, \& Wilkinson 2001;
Ofek et al. 2001;
Keeton \& Madau 2001; Phillips et al. 2001b)

The incidence of large-separation gravitational lensing
depends on several
factors, including
the cosmology, the present-day cluster mass function and its
redshift evolution, and most importantly,
the mass profile and the substructure within galaxy clusters
(e.g., Bartelmann, Steinmetz, \& Weiss 1995).
There are several approaches for predicting the optical depth
to large-separation lensing
in a given model scenario:

(i) Ray tracing through $N$-body simulations (e.g., 
Cen et al. 1994; Wambsganss et al. 1995;
Wambsganss, Cen, \& Ostriker 1998).
This accounts for the
complicated large-scale, multi-plane lens structure,
and for the substructure in clusters of galaxies.
Wambsganss et al. (1995; 1998) predict an optical depth for
lensing with $\Delta\theta>5''$ of about $3\times10^{-3}$,
for a $z\approx3$ source in an $\Omega_{0}=1$ Cold Dark Matter (CDM) cosmology.
The problem of substructure,
in particular, was analysed
by Bartelmann \& Weiss (1994), and Bartelmann et al. (1995).
They found that numerically modeled clusters with substructure are
about two orders of magnitude more efficient in producing
highly magnified images (e.g., long arcs)
than spherically symmetric clusters.
%\newline

(ii) Taking the observed cluster mass function and
adopting a smooth and symmetric
mass profile to calculate the optical depth
(e.g., Maoz et al. 1997; Ofek et al. 2001).
Maoz et al. (1997) used the NFW profile
(Navarro, Frenk, \& White 1996, 1997)
to show that the cluster lensing
optical depth can be in the range
$\tau(\Delta\theta>5'')\approx10^{-5}$ to $10^{-2}$,
where $\tau$ depends
mostly on the adopted mass-concentration.
In Ofek et al. 2001 (hereafter Paper~I), we estimated the optical depth for
singular isothermal sphere (SIS) lenses
using the non-evolving Girardi et al. (1998) cluster
mass function
and found
$\tau(\Delta\theta>5'')\approx3\times10^{-4}$,
for an $\Omega_{0}=0.3$, $\Omega_{\Lambda}=0.7$ cosmology and
a source redshift of $z_{source}=1$.

(iii) Modifying approach (ii) by
assuming the 
cluster mass function is given by
the Press-Schechter formalism (Press \& Schechter 1974),
that by itself depends on cosmology.
For example,
Li \& Ostriker (2002) have followed this approach, and assumed
a generalized NFW density profile and $\bar z_{source}=1$.
Their predictions for the optical depth span values in an even
larger range,
$7\times10^{-7} \ltorder \tau(\Delta\theta>5'') \ltorder 10^{-3}$,
depending in part on the details of the  adopted mass profile.
The
impact of the cosmological model on the
cross sections may seem surprisingly large,
about one or two orders of magnitudes.
This dependence enters
through the differing
redshift
evolutions of the
cluster
mass function.
Other examples of the Press-Schechter approach to cluster lensing statistics
are Narayan \& White (1988), Flores \& Primack (1996), 
Wyithe, Turner, \& Spergel (2001), and Sarbu, Rusin, \& Ma (2001).

In light of these wide-ranging predictions,
empirical constraints seem in order.
Without any
confirmed large-separation quasar lenses, we can only place
an upper limit on the cross section which, in turn, can
reject part of the mass-profile and cosmology parameter space.

In Paper~I we described
a survey for large separation gravitational lensing among
quasars drawn from the FIRST radio catalog.
In that survey, we selected radio sources from the FIRST catalogue
(1999, July 21 version;
Becker, White, \& Helfand 1994; White et al. 1997)
with
optical counterparts in the APM catalog (McMahon, \& Irwin 1992),
with positional coincidence better than
$2.''5$ in both the $O$ and $E$ bands, with colour
index $O-E<2$~mag, and that are point-like
on at least
one of the
$O$ or $E$ POSS-I/UKST plates,
as determined by
the APM classification algorithm.
There are $12,576$ quasar candidates passing
the above criteria.
Based on the initial results from the
FIRST Bright Quasar Survey (FBQS; Gregg et al. 1996; White et al. 2000),
we estimated in Paper~I that there are about $9100$ quasars in our sample
out of the $12,576$ candidates.

However, with the more recently published results of the FBQS 
(Becker et al. 2001), which extends the
spectroscopic follow-up
from 17.8 to 19~mag,
we can revise
the fraction of
quasars among the sources fainter
than $E=17.8$~mag
in our sample from $90\%$ to $80\%$, or about $8000$ quasars
instead of $9100$ quasars.
Specifically, Equation~1 in Paper~I can be revised to:
\begin{equation}
F_{qso} = \left\{ \begin{array}{ll}
          -0.9389 + 0.046E + 0.0027E^{2}, & \mbox{$E<18$} \\
          0.8,                            & \mbox{$E\geq 18$}
          \end{array} \right.
\label{Frac_qso}
\end{equation}

Follow-up
observations of all
pairs of radio-optical
quasar candidates with
$5''<\Delta\theta<30''$
showed that none of the 15 candidate pairs are
lensed quasars (for details see Paper~I).
For $8000$ quasars in the sample,
this implies an upper limit of $3.7\times10^{-4}$
% 2.995./8000=3.744e-4
($95\%$ Poisson statistics confidence level; CL)
on the lensing fraction in this survey.

However, in order to constrain the
mass function and mass profile of clusters from these results,
some further properties of our survey need to be known:
(i) the double flux magnification bias of our survey, resulting
from the radio and optical flux limits involved in the sample selection.
This magnification bias reflects the over- (or under)
representation of lensed objects in our survey,
compared to purely geometrical cross-section calculations; and
(ii) the redshift distribution of quasars
(i.e., the potential sources) in our sample,
which determines the effective pathlength of our lens search.

The purpose of the present paper is to improve 
those two important constraints
and provide a more reliable upper limit on the
observed optical depth to large separation lensing.
 
\section{The survey double flux magnification bias}
\label{Sec_mag_bias}

As described in Paper~I, the magnification bias depends
strongly on the
faint-end number counts (in the radio and optical) of quasars 
%that have selection criteria similar to
with similar selection criteria to those
used in our survey.
However, there are presently no large-area very deep radio-optical
surveys.
The best available data set is the combination of
radio and optical surveys in the
Hubble Deep Field North (HDF-N) region:
a deep $1.4$~GHz radio survey
conducted with the VLA by Richards (2000).
This survey has a flux limit ($5\sigma$)
of $40$~$\mu$Jy in the central $8'$,
degrading to about $300$~$\mu$Jy at $18'$ from the field centre.
In the optical, a $1$~deg$^{2}$ region around the HDF-N
was searched for quasars down to $B=22$~mag
using multi-colour selection (Liu et al. 1999; Vanden Berk et al. 2000).
Liu et al. required
somewhat more restrictive color criteria than ours:
$B-R<0.8$,  or $U-B\leq-0.4$ {\it and} $B-R\leq1.1$.
Impey \& Petry (2001) matched the Richards (2000) radio sources with
the optically detected quasars in this field, and
identified three radio-optical quasars within the colour range,
$0.46<B-R<0.75$, that match our survey criteria.
With sufficient
lensing magnification these quasars
would have been detected by our survey.
These data
on the HDF-N constitute faint-end counts that can be
combined with our FIRST-APM candidates,
to find the double flux magnification bias.

\begin{figure}
\centerline{\epsfxsize=85mm\epsfbox{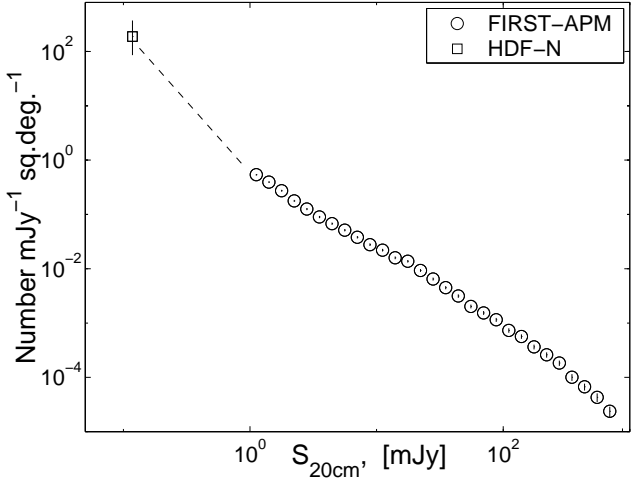}}
\caption {Differential number counts in the $1.4$~GHz-band
for our sample quasars (circles), and one square representing
the deep radio-optical search for quasars described above.
The dashed line is the interpolated number-counts in
the $0.1$-$1$~mJy range.}
\label{NumCount_20cm_100microJy}
\end{figure}

\begin{figure}
\centerline{\epsfxsize=85mm\epsfbox{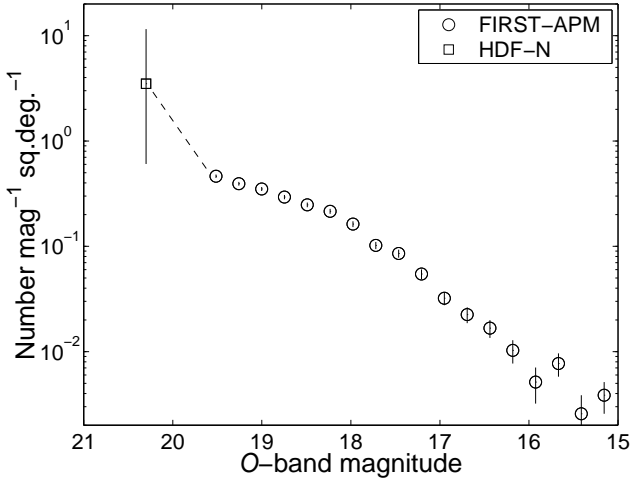}}
\caption {Same as Figure~\ref{NumCount_20cm_100microJy},
but for the $O$-band}
\label{NumCount_B_21}
\end{figure}

The HDF implies radio counts of
$190^{+180}_{-100}$
radio-optical
sources per deg$^{2}$ per mJy, at $0.12$~mJy in the 
$1.4$~GHz-band,
and
optical counts of
$3.5^{+8.1}_{-2.9}$ radio-optical sources per deg$^{2}$
per mag at $O=20.3$ magnitude.
The new radio number-count point includes three radio-optical quasars
with optical counterpart brighter than $B=22$,
and the optical point is based on one faint
(fainter than the APM $100\%$ completeness limit of $19.75$;
Caretta et al. 2000)
radio-optical quasar
with radio counterpart brighter than the Richards (2000) survey limit, as
given above.
Figure~\ref{NumCount_20cm_100microJy}
shows the differential number counts in the $1.4$~GHz-band
for the Paper I sample quasars (circles), and for
the deep radio-optical search for quasars described above (single square).
The dashed line is the interpolated number-counts in
the $0.1$-$1$~mJy range.
Figure~\ref{NumCount_B_21} is
the same as Figure~\ref{NumCount_20cm_100microJy},
but for the optical $O$-band.

Although it is reasonable to assume that there is a large
number of radio-faint (fainter than $0.1$~mJy)
quasars constituting the radio-quiet
quasar population,
we only have direct observational constraints on the
number counts down to $0.1$~mJy
(Figures~\ref{NumCount_20cm_100microJy}-\ref{NumCount_B_21}).
Therefore, our estimate for the double-flux magnification bias
will only be a lower limit.
Furthermore, due to the small area of the HDF-N deep radio observations,
the number counts
at the intermediate flux levels (between FIRST and HDF-N)
are not measured directly, and we simply
interpolate the number count distribution over these gaps.

Drawing on this new information we
have calculated the double flux magnification bias
in the FIRST-APM sample
through a Monte-Carlo simulation, as follows.
\newline
%(i)
We constructed a list of radio-optical quasars
that mimics the observed number count distribution of the
$O$-magnitude and of the $1.4$~GHz flux shown
in Figures~\ref{NumCount_20cm_100microJy}-\ref{NumCount_B_21}.
For quasars above the FIRST-APM flux limits we drew at random
objects from our list of quasar candidates,
corrected for the
fraction $F_{qso}$ of these objects
that are quasars,
%(ii)
and for the incompleteness
of the APM survey
(Caretta et al. 2000), by replicating
objects in the magnitude-range $19.5$-$21.5$,
until
we had a list of about $8000$ radio-optical sources.
%(iii)
At the faint end
we added objects with
radio flux and optical magnitude
drawing from a
flux probability distribution
obtained by interpolating the number counts
in Figure~\ref{NumCount_20cm_100microJy}.
For each of these objects we then drew 
an $O$-magnitude from the distribution given
in Figure~\ref{NumCount_B_21}.
Note that the last step
%(and step-i)
assumes that there is no
correlation between the optical and radio properties
of quasars,
which is true at least
for objects brighter than $1$~mJy and $O\sim21$ (Paper~I).
The number of randomly selected objects in this step
was chosen by scaling the number of HDF-N faint sources by
 the ratio of the total area of the FIRST-APM survey
to the area of the HDF-N faint-end number-count survey
described above.
\newline
%(iv)
With the list described above,
we are now in a position to estimate
the double-flux magnification bias,
assuming a SIS lens mass profile.
Our FIRST-APM survey will detect a lensed quasar as such
only if {\it both} lensed images (bright and faint) are
brighter than the radio $1$~mJy flux limit, {\it and}
are brighter than
the $O$-magnitude limit of $21.2$,
which is the weighted average magnitude
limit of the APM (Caretta et al. 2000).
The magnification/de-magnification of the faint image, $A_{-}$,
in a SIS lens is given by: $\beta = 1/(1+A_{-})$,
where $\beta$ is the impact parameter in units of the
Einstein ring radius.
By comparing the magnitude and flux
of each object in the Monte-Carlo list
with the survey's flux limits,
we calculated for each object, $i$,
the minimum magnification needed (or maximum de-magnification allowed), $
A_{i,min}$,
and the corresponding $\beta_{i,max}$,
needed for the detection of an image in
both the radio and optical bands
in our survey (i.e., the smaller among the optical and
radio values of $\beta_{i,max}$ was chosen).

The cross-section for strong lensing is
defined as the area with $\beta\leq1$,
and the probability for a given impact parameter
$\beta$ scales like the area, $\beta^{2}$.
To estimate the double-flux magnification bias, $B$,
we summed $\beta_{i,max}^{2}$ for all the objects in our Monte-Carlo list
and divided the results by the number of quasars in our
sample, i.e., $8000$.
The resulting ratio reflects the over- (or under) representation
of lensed objects in our survey, compared to a purely
geometrical cross-section calculation.

Using the above method, we obtain a lower
limit of $B\gtorder1.1$. This limit changes to $0.9$
if we take into account the $1\sigma$
Poisson statistics (lower confidence interval of the number counts),
and to $1.3$ if we assume that
the optical searches for quasars described above are $50\%$
incomplete (see Impey \& Petry 2001) due to the different
colour selection process used by Liu et al. (1999)
and Vanden Berk et al. (2000).
%---
The $B>0.9$ limit we give above assumes that the number-counts
are almost flat from $O\sim19$ to $O\sim20$,
and then go to zero.
%---
The actual value of $B$ could be
larger due to the probable existence of a yet-fainter
radio-optical quasar population.
Hence, we consider $B\geq1.1$ to be a conservative limit.
%However,
%it is unlikely that $B$ is much larger,
%as very faint quasars will be at larger distances
%and redder, and therefore they will not pass our
%survey colour criterion.
%
For comparison, in Paper~I we roughly estimated $B$ by
assuming that the product of the optical and radio number-flux relations
is a  power-law with a cutoff at a flux
that is $0.5-4$ orders of magnitude lower than our
survey's flux/magnitude limits.
Based on this, we estimated $B$ to be in the range of $1-40$.
Since the HDF-N data
extend the depth of our optical source counts by about 1~mag, 
the lower limit
on $B$ found in this paper is near
the low-end estimate of Paper~I, but is now on a firmer observational basis.
The present method we have used to calculate the bias is superior to the
Borgeest, Linde, \& Refsdal (1991) approach we used in Paper~I,
since it takes into account the actual shapes of the separate optical
and radio number count distributions instead of
replacing them by single power-law distributions.

\section{Redshift Distribution of the FIRST/APM Sources}
\label{Sec_redshift_dist}
From the first major public release of the
Sloan Digital Sky Survey - Early Data Release
(SDSS-EDR; York et al. 2000; Stoughton et al. 2002),
we constructed a sample of
$443$
point-like SDSS objects
that: were selected as quasar-candidates
in the SDSS-EDR based on their colors;
are found within
$1.''2$
from a FIRST radio source; and are brighter
than $g'=21$~mag.
Their magnitudes were corrected for Galactic extinction (Schlegel, Finkbeiner, \& Davis 1998).
All but two sources in this list have $g'-r'<2$~mag, so
they are good representatives of our survey.
Based on the SDSS quasar selection criteria
(e.g., Newberg \& Yanny 1997; York et al. 2000; Schneider et al. 2002), and on the FBQS
(Gregg et al. 1996; Helfand et al. 1998; Becker et al. 1998;
White et al. 2000; Becker et al. 2001),
we know that the fraction of quasars in this sample is $\gtorder75\%$.

\begin{figure}
\centerline{\epsfxsize=85mm\epsfbox{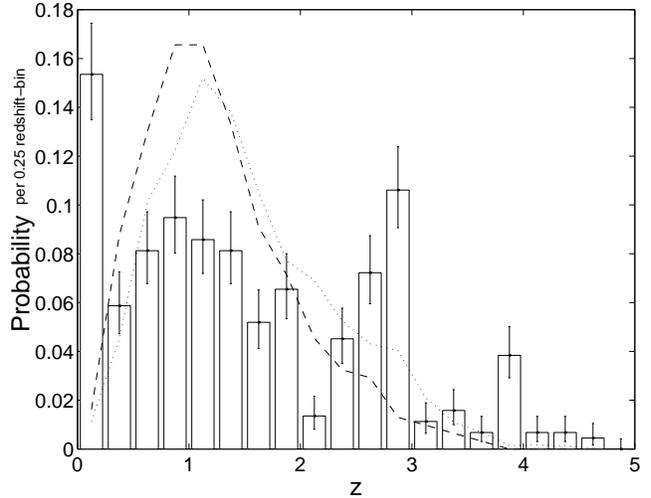}}
\caption {The redshift distribution of FIRST quasar
 candidates down to $g'=21$~mag (bars). 
The redshift distribution was calculated using a 
simple colour-redshift relation.
The superposed vertical-lines on the bars, are the Poisson
error-bars due to the finite sample size.
The dashed line is the redshift distribution of
FBQS faint-extension ($E<19$~mag) quasars.
The dotted line is the same distribution, but
convolved with the
photometric redshift scatter, as a function of redshift.}
\label{ObsPredConvErr_photzErr_FFQS_EDR}
\end{figure}

From the quasar redshift-colour relation,
given by Richards et al. (2001a, Table $3$),
we estimated the photometric redshifts
for our
$443$-object sample,
by a simple $\chi^{2}$ minimization (Richards et al. 2001b)
between the predicted and the observed  SDSS colours.
%---
The photometric-redshift distribution of these objects is shown as bars
in Figure~\ref{ObsPredConvErr_photzErr_FFQS_EDR}.
%The superposed vertical lines on the bars are the Poisson
%errors due to the finite sample size.
%---
The dashed line shows the redshift distribution of quasars in the
FBQS faint extension ($E<19$~mag; Becker et al. 2001).
%---
The photometric-redshift distribution is not changed considerably
if only FIRST sources with SDSS
counterparts brighter than $g'=19$~mag are taken.
%---
However, our photometric redshift estimate of quasars has a
considerable scatter: $30\%$ of the redshifts
have errors, $\Delta{z}>0.25$.
%---
%The scatter in the photometric redshift is a function of redshift,
%i.e., in some regions of the quasar colour-space, there
%is a redshift-colour degeneracy, resulting in a significant fraction
%of outliers.
%In order
%---
To estimate the effect of this scatter on the
photometric redshift distribution,
we have used the $898$ quasars with known redshifts and SDSS colours
given by Richards et al. (2001a)
to calculate the
distribution of differences between our redshift prediction
and the observed redshifts in each $0.25$ redshift bin.
We convolved this scatter distribution as a function of redshift with
the FBQS faint-extension redshift distribution
($E<19$~mag; Becker et al. 2001)
and the result is shown as the dotted line in
Figure~\ref{ObsPredConvErr_photzErr_FFQS_EDR}.
Ideally, we would expect that the photometric redshift distribution (bars) 
be similar to the redshift distribution of FBQS quasars
convolved with the photometric-redshift scatter (dotted line).
However,
our photometric-redshift distribution has
an excess of quasars at $z\sim3$, relative to the
FBQS faint-extension redshift distribution
%---
%convolved with the
%photometric-redshift scatter.
%---
The median redshift for the sample of FBQS faint-extension
quasars is $\bar z=1.2$,
and the median redshift we get from the photometric redshift
down to limiting magnitude $g'=21$ is $\bar z=1.4$.
There are two possibilities to explain the discrepancy:
(i) only a fraction ($\gtorder75\%$) of the sources in our sample are real quasars.
Therefore, the non-quasar objects could contaminate the redshift distribution
in some systematic way;
(ii) the photometric-redshift procedure of Richards et al. (2001b) was
developed based mostly on radio-quiet quasars, but
our survey contains mainly radio-loud quasars.
It is known that there are some optical spectral differences between radio-loud
and radio-quiet quasars (e.g., Brotherton et al. 2001).
Such differences could bias the Richards et al. (2001b)
photometric-redshift procedure.
% For example, radio-loud quasars are redder...
Therefore we
%---
%treat the redshift distributions
%in Figure~\ref{ObsPredConvErr_photzErr_FFQS_EDR} with caution,
%and
%---
adopt the FBQS faint-extension redshift distribution
for our calculations. This is a conservative choice, as the lower
median redshift implies a smaller path length for lensing.
%---
% and hence
%a lower predicted lensing fraction.
%---

\section{Discussion}
\label{discussion}

In the following sections we use these two new pieces of information,
$B\geq1.1$ and the redshift distributions, to estimate
the optical-depth to lensing in a simple SIS model,
and to compare our results with $\Omega_{0}=1$
CDM $N$-body ray-tracing simulations.

\subsection{A SIS model}
\label{SSM}
To get some simple estimates of the lensing probability
in our survey,
we calculated the cross section as a function of redshift,
assuming SIS lenses and
the Girardi et al. (1998) mass function
of groups and clusters with no evolution.
We found in Paper~I that the effective dimensionless density of lenses
(as defined by Turner et al. 1984) is,
$F=16\pi^{3}n_{0} \left(\frac{c}{H_{0}} \right)^{3} \left( \frac{\sigma_{\parallel}}{c} \right)^{4} = 0.0156_{-0.0067}^{+0.012}$,
where $c$ is the speed of light, $H_{0}$ the Hubble constant,
$\sigma_{\parallel}$ the line-of-sight velocity dispersion,
and $n_{0}$ is the comoving number-density of lenses.
The errors in $F$ were obtained from Monte-Carlo simulations
using the Girardi et al. (1998) mass function
parameters and their errors (for details see Paper~I).
In Paper~I we assumed a single representative redshift of $z=1$ for all
the quasars.
With our new results presented above,
we re-calculate the total optical-depth for lensing,
by integrating the optical-depth over redshift,
with the weighting function given by the
FBQS faint-extension redshift probability density distribution
(dashed-line in Figure~\ref{ObsPredConvErr_photzErr_FFQS_EDR}).
The optical depth $\tau_{z}$, to a source at redshift $z$, is given by:
\begin{equation}
\tau_{z} = \int_{0}^{z}{ F(1+z_{l})^{3} \left( \frac{D_{l}D_{ls}}{D_{s}} \right)^{2} \frac{1}{R_{0}} \frac{cdt}{dz_{l}}dz_{l} },
\label{tauz}
\end{equation}
where $D_{l}$, $D_{ls}$, $D_{s}$ are the angular-diameter distances
for the observer-lens, lens-source and observer-source, respectively,
$z_{l}$ is the lens redshift, and
$R_{0}\equiv c/H_{0}$.
The quantity $cdt/dz_{l}$ is calculated in the
Friedmann-Lema\^{i}tre-Robertson-Walker geometry
(e.g., Fukugita et al. 1992).
The total weighted optical depth is given by:
\begin{equation}
\tau = \int_{0}^{\infty}{P(z)\tau_{z}dz}
\label{tottau}
\end{equation}
where $P(z)$ is the
normalized weighting function given by
the FBQS quasar redshift distribution.
We find
\begin{equation}
\tau = \left\{ \begin{array}{ll}
          4.6_{-2.0}^{+3.5}\times10^{-4}, & \mbox{$\Omega_{0}=0.3; \Omega_{\Lambda}=0.7$}   \\
          2.4_{-1.0}^{+1.8}\times10^{-4}, & \mbox{$\Omega_{0}=0.3; \Omega_{\Lambda}=0.0$}   \\
          1.7_{-0.7}^{+1.3}\times10^{-4}, & \mbox{$\Omega_{0}=1.0; \Omega_{\Lambda}=0.0$} , \\
          \end{array} \right.
\label{tau_final}
\end{equation}
where the errors are the $1\sigma$ errors due to the uncertainty in $F$.
This is $\sim1.7$ times greater than estimated in Paper I, where $z=1$
was assumed for all sources.
Using the photometric-redshift distribution from
Figure~\ref{ObsPredConvErr_photzErr_FFQS_EDR} (bars) instead of the
FBQS faint-extension redshift distribution,
the predicted $\tau$ would further increase
by a factor of about $1.6$.
The difference is due to the excess of quasars
with $z\sim3$ in our photometric-redshift distribution,
combined with the fact that the optical-depth increases with redshift.
% 1.5 to 1.9 for different cosmologies...
We conservatively adopt the optical-depth in Equation~\ref{tau_final},
as a lower limit on $\tau$.
These optical-depths need to be multiplied by the completeness
factor that reflects the range of separations probed.
As shown in Paper~I, this completeness factor is about $0.66$
for a SIS model.
Finally, the lower limit on the magnification ($B\geq1.1$)
determines the expected probability for lensed quasars in our sample:
$P_{lens}\geq4.6_{-2.0}^{+3.5}\times10^{-4}\times0.66\times1.1=3.3_{-1.5}^{+2.5}\times10^{-4}$
(for $\Omega_{0}=0.3$, $\Omega_{\Lambda}=0.7$).

The observational 
upper bound on the lensed fraction of $3.7\times10^{-4}$ ($95\%$ CL)
from our survey (Paper~I, and small revisions in this paper),
is therefore consistent with this simple
model ($P_{lens}$). Since the ($\Omega_{0}=0.3, \Omega_{\Lambda}=0.0$)
and ($\Omega_{0}=1.0, \Omega_{\Lambda}=0.0$) models predict a lower
optical depths, they too are consistent with the data.
Within this simple model,
for an $\Omega_{0}=0.3$, $\Omega_{\Lambda}=0.7$ cosmology
and a non-evolving Girardi et al. (1998) mass function,
we can in turn set an upper limit to $F<0.018$ with a $95\%$ CL.
%---
Models incorporating evolution of the mass function will reduce
the optical-depth, therefore increasing the upper limit on $F$.
%---

\subsection{Comparison with ray-tracing simulations}
\label{CRT}
Wambsganss et al. (1995, 1998) performed ray tracing
for $0<z<3$
through a three-dimensional mass distribution
obtained from an $\Omega_{0}=1$ COBE normalized CDM $N$-body simulation.
They found that the optical depth for strong lensing
(i.e., for forming multiple images)
with image separations $>5''$ of sources at $z\sim3$
is $3\times10^{-3}$.
Our observations can test this model.
Figure~2 in Wambsganss et al. (1995) describes the
splitting probability for image pairs with separations
greater than $5''$ and magnitude differences
smaller than $1.5$,
as a function of the source redshift.
To obtain the expected optical-depth for our survey,
we integrate their splitting probability with a
weight function given
by the FBQS faint-extension source redshift distribution
presented in Figure~\ref{ObsPredConvErr_photzErr_FFQS_EDR}.
We obtain $\tau=1.14\times10^{-3}$
(or $\tau=1.5\times10^{-3}$ using the photometric-redshift distribution).
This optical depth
needs to be multiplied by the completeness factor
for the range of separations probed and the magnification bias
of our survey.
Wambsganss et al. (1995) also show (their Figure~3) the
strong-lensing probability as a function of the image
separation for various source redshifts.
From this figure we estimate the incompleteness due to
the separation range probed in our survey,
$P(\Delta>30'')/P(\Delta>5'')\sim 0.25$
(This separation incompleteness is actually slightly
redshift dependent, and changes from $\sim30\%$
at $z_{s}=1$ to $\sim20\%$ at $z_{s}=3$).
Taking together $\tau$, the separation incompleteness
and the magnification bias we have found here,
implies a
probability for lensing in our survey
$P_{CDM}\sim1.14\times10^{-3}\times0.75\times1.1=9.4\times10^{-4}$
based on the Wambsganss et al. simulations.

We note that the magnification bias we found in this paper is
based on a SIS flux ratio distribution
- an assumption that is not necessarily compatible with the
Wambsganss et al. (1995, 1998) simulations.
However,
Figure~5 in Wambsganss et al. (1998)
shows that the magnification probability
found in their simulations is not
very different from the $P(A)\sim A^{-3}$
expected from the SIS model we assumed in our
magnification-bias calculation.
Based on the calculation described above,
7.5 lenses are expected in our survey
for an $\Omega_{0}=1$ CDM model.
We can therefore reject the
Wambsganss et al. (1995, 1998) model at $99.9\%$ CL.
We note that, since Figure~2 in Wambsganss et al. (1995)
gives the splitting angle probability for image flux ratios
smaller than 1.5~mag, our conclusion is conservative.
%---
Note, that the simulations of Wambsganss et al. use
the COBE normalization, $\sigma_{8}=1.05$,
which predicts an excess of high-mass clusters at the present
epoch, and slower evolution of the mass function.
In this sense, our results do not reject the $\Omega_{0}=1$ model
{\it per se}, but the combination of the cosmology and the high normalization.
%---

\vspace{0.5cm}

The results presented here 
constitute one of the tightest and well-defined
limits on large separation lensing to date.
Clearly, we have now reached a level where the
non-detections are becoming interesting cosmological constraints.
It is probable that actual cases of large-separation gravitationally
lensed quasars
will be found soon using one of the large area surveys
(e.g., SDSS, 2DF; see Croom et al. 1998).
However, in order to use such large separation lenses
to constrain the cluster mass function and mass profile parameter space,
we need to: (i) understand the selection effects and effectiveness of these surveys for large separation lensing;
(ii) have a realistic estimate of the cross section for large separation
lensing, taking into account the substructure of clusters.
These points will be addressed in future papers.

To summarize our main results, we have derived a lower limit to
the magnification bias in our survey, $B\geq1.1$, and we have also found that
the median photometric-redshift of quasars in
our sample is $\bar z_{source}\sim1.4$.
Using the redshift distribution from the FBQS faint extension,
we determine that $F<0.018$ with $95\%$ CL
(assuming a SIS model and an $\Omega_{0}=0.3$, $\Omega_{\Lambda}=0.7$
cosmology). Our non-detection of lensed FIRST quasars is consistent
with expectations, if clusters can be represented by a non-evolving
population of SIS masses with the mass function of the observed cluster
population. If so, moderately larger surveys will discover the first 
examples of large separation quasar lensing. Our survey already has
the ability to reject some models with concrete predictions, namely 
the $\Omega_{0}=1$ COBE-normalized CDM model, whose excess of power on 
large scales is well known.

\section*{ACKNOWLEDGMENTS}

We thank  Eric Richards for sending us 
a digital version of his deep radio catalogue of the HDF region,
and an anonymous referee for useful comments.
EOO wishes to thank 
the Max-Planck-Institut f\"{u}r Astronomie
for its hospitality and financial support;
the Deutscher Akademischer Austauschdienst
for financial support; and
Orly Gnat and Avishay Gal-Yam for fruitful discussions.
This research has made use of the Sloan Digital Sky Survey.
Funding for the creation and distribution of the SDSS Archive has been
provided by the Alfred P. Sloan Foundation, the Participating
Institutions, the National Aeronautics and Space Administration,
the National Science Foundation,
the U.S. Department of Energy,
the Japanese Monbukagakusho,
and the Max Planck Society.
The SDSS Web site is http://www.sdss.org/.

\end{document}